\documentclass[twocolumn,prl,aps,a4,floatfix,amsmath,amssymb]{revtex4-1}

\usepackage{amssymb}
\usepackage{epsfig}
\usepackage{amsmath}
\usepackage{amsfonts}
\usepackage{graphicx}
\graphicspath{{figure file fold path}}
\usepackage[dvipsnames,usenames]{color}

\begin{document}
\title{
Analytical vs. Numerical Langevin Description of Noise in Small Lasers
}
\author{G.L. Lippi}
\affiliation{Universit\'e C\^ote d'Azur, Institut de Physique de Nice (INPHYNI), CNRS UMR 7010, Nice, France}
\email{gian-luca.lippi@inphyni.cnrs.fr}
\author{J. M\o rk}
\affiliation{DTU Fotonik, Technical University of Denmark, 2800 Kgs. Lyngby, Denmark}
\author{G.P. Puccioni}
\affiliation{Istituto dei Sistemi Complessi, CNR, Via Madonna del Piano 10,
I-50019 Sesto Fiorentino, Italy}

\date{\today}

\begin{abstract}
We compare the analytical  and numerical predictions of noise in nano- and microcavity lasers obtained from a rate equation model with stochastic Langevin noise.  Strong discrepancies are found between the two approaches and these are critically analyzed and explained on the basis of general considerations and through the comparison to the numerical predictions of a Stochastic Laser Simulator.  While the analytical calculations give reliable predictions, the numerical results are entirely incorrect thus unsuitable for predicting the dynamics and statistical properties of small lasers.
\end{abstract}


\maketitle

Since their conception~\cite{Yokoyama1989}, nanolasers have been a source of questions concerning the fundamentals of laser physics.  One crucial point centers around their dynamical and statistical description as the $\beta$--parameter (fraction of spontaneous emission coupled into the lasing mode) increases.  Indeed, as the modal volume is reduced, corresponding to the transition between the thermodynamic limit of $\beta \ll 1$ and the thresholdless case of $\beta \approxeq 1$, fluctuations play an ever increasing role~\cite{Rice1994} thereby calling into question the semiclassical description.  The discreteness of the processes (exchange of photons and carriers in integer numbers) become also of paramount importance~\cite{Lebreton2013}, suggesting that the Rate Equations (REs)~\cite{Coldren2012} no longer be usable for meaningful predictions.  However, REs have been proven time and again to provide meaningful estimates of dynamics and noise~\cite{Gies2007,Chow2014,Ding2015,Romeira2018,Mork2018}, i.e., when comparing to experiments~\cite{vanDruten2000,Ota2017,Kreinberg2017}, extracting experimental laser parameters~\cite{Strauf2006}, or estimating possible data transmission rates~\cite{Moelbjerg2013}.
The purpose of the present contribution is to point put out important discrepancies between analytical and numerical predictions from the REs and explain the origin of this unexpected difference.

The fundamental hypotheses on which a Langevin description is based rest on the one hand on the smallness of the perturbation applied to the macroscopic, deterministic variable, on the other hand on the fast time scale over which the perturbations act~\cite{Langevin1908}, compared to the intrinsic dynamics.  The first condition is difficult to fulfill in a nanolaser since the photon number at threshold scales as $\beta^{-\frac{1}{2}}$~\cite{Rice1994}, thus $\Delta n \ll \langle n \rangle$ photons requires at least a ``borderline'' nanolaser ($\beta = O(10^{-2}) \Rightarrow \langle n \rangle = 10$ for $\Delta n = 1$); however, mesoscale devices would qualify~\cite{Wang2018}.  Before analyzing the second condition (time scale), important also for the numerical integration of the model equations, we are going to introduce the REs that we use.

A standard form of the REs, with the inclusion of the usual Langevin Noise terms (hereafter RELN), reads~\cite{Coldren2012}:
\begin{eqnarray}
\label{fre1}
\dot{n} & =  & -\Gamma_c n + \beta \gamma N (n+1) + F_{ph}(t)\, , \\
\label{fre2}
\dot{N} & = & R - \beta \gamma N n - \gamma N + F_{c}(t)\, , 
\end{eqnarray}
where $n$ and $N$ represent the photon and carrier number (or population inversion), respectively, $\Gamma_c$ and $\gamma$ are the relaxation rates for the intracavity
photons and for the population inversion, respectively, and $R$ is the pump
rate.  This set of RELNs includes the average contribution of the spontaneous emission to the number of coherent photons in the cavity mode through the term $\beta \gamma N$.  The Langevin noise terms, reworked from~\cite{Coldren2012}, read:
\begin{eqnarray}
\label{photnoise}
F_{ph} & = &  \sqrt{2} \left( \sqrt{F_{nn}} G_a(0,1) + \sqrt{F_{nN}} G_b(0,1) \right) \, , \\
\label{popnoise}
F_c & = &  \sqrt{2} \left( \sqrt{F_{Nn}} G_c(0,1) + \sqrt{F_{NN}} G_d(0,1) \right) \, .
\end{eqnarray}
$G_j(0,1)$'s are independent Gaussian processes ($G_j \ne G_k$, $j \ne k$) with zero average and unity variance and
\begin{eqnarray}
F_{nn} & = & 2 \beta \gamma \overline{N} ( \overline{n} + 1 )\\
F_{nN} = F_{Nn} & = & - \beta \gamma \overline{N} \left( \overline{n} + 1 \right)\\
F_{NN} & = & R + \gamma \overline{N} + \beta \gamma \overline{N} \overline{n} 
\end{eqnarray}
are the $\delta$-correlated noise contributions~\cite{Coldren2012} computed following the McCumber approach~\cite{McCumber1966}.

The steady state solution of eqs.~(\ref{fre1},\ref{fre2}) is:
\begin{eqnarray}
\label{nss}
\overline{n} =  \left\{ \left(\frac{C-1}{2}\right) + \sqrt{ \left(\frac{C-1}{2}\right)^2 + \beta C} \right\} \beta^{-1} \, , \\
\label{Nss}
\overline{N} = \frac{\Gamma_c}{\beta \gamma} \frac{C}{1 + \beta \overline{n}}\, , \quad 
C = \frac{R}{R_{th}}\, , \quad R_{th} = \frac{\Gamma_c}{\beta} \, ,
\end{eqnarray}
which give the threshold ($C=1$) values:
\begin{eqnarray}
\label{thss}
\overline{n}_{th} = \frac{1}{\sqrt{\beta}} \, , & \quad &
\overline{N}_{th} = \frac{\Gamma_c}{\gamma_{\parallel} \beta} \frac{1}{1+ \sqrt{\beta}} \, .
\end{eqnarray}

Two time scales emerge from the REs:  $\gamma$ (relaxation constant for the carriers) and $\Gamma_c$ (inverse photon lifetime in the laser cavity), where the class B laser condition~\cite{Tredicce1985} $\gamma \ll \Gamma_c$ holds for standard semiconductor lasers.  For the following, it is important to remark that $\Gamma_c$ corresponds to the fastest dynamical time scale describing the laser (after the adiabatic elimination of the medium's polarization), while $\gamma$ represents the rate of spontaneous relaxation of the excited state (i.e., electron-hole recombination).

Photon emissions occur independently of each other and therefore follow Poisson statistics ($\mathcal{P}(x)$, where $x$ is the average number of events in a given time interval).  The Langevin noise description involves Gaussian processes, eqs.~(\ref{photnoise},\ref{popnoise}), and can be considered a satisfactory approximation ($\mathcal{G}(x) \approx \mathcal{P}(x)$) when $x$ is large~\cite{Walck2007} (in practice $x > 10$).  Identifying $x$, in eq.~(\ref{fre1}), with the spontaneous emission or, in turn, the stimulated process, we arrive at the following conditions
\begin{eqnarray}
\label{spontcond}
\gamma \beta \overline{N}_{th} \Delta t_{sp} \gg 1 \, , \\
\label{stimcond}
\gamma \beta \overline{N}_{th} \overline{n}_{th} \Delta t_{st} \gg 1 \, , 
\end{eqnarray}
which provide the restrictions on the time intervals $\Delta t_{sp}$ (eq.~(\ref{spontcond})) and $\Delta t_{st}$ (eq.~(\ref{stimcond})) fixing the shortest observation timescales to satisfy the replacement of Poisson with Gaussian statistics.  

The laser threshold, as a critical point, is the most suitable choice for testing the validity of the Gaussian approximation.  Substitution of the steady-state expressions, eqs.~(\ref{thss}), into eqs.~(\ref{spontcond},\ref{stimcond}),  yields:
\begin{eqnarray}
\label{condsp}
\Delta t_{sp} & \gg & \frac{1 + \sqrt{\beta}}{\Gamma_c} > \frac{1}{\Gamma_c} \, , \\
\label{condst}
\Delta t_{st} & \gg & \frac{\beta + \sqrt{\beta}}{\Gamma_c} > \frac{\sqrt{\beta}}{\Gamma_c} \, .
\end{eqnarray}
Since $\Delta t_{sp} > \Delta t_{st}$, it is sufficient to choose a numerical time step $\Delta t_s \ge \Delta t_{sp}$ to warrant the replacement of Poisson with Gaussian statistics.

The Euler-Maruyama scheme is the simplest, yet very effective integration method that can be applied to the RELNs~\cite{Lippi2018}.  As in all numerical schemes, its time step, $\Delta t_s$, must be much smaller than the shortest time constant present in the system, i.e., 
\begin{eqnarray}
\label{timestepcond}
\Delta t_s \ll \Gamma_c^{-1}
\end{eqnarray}
in our RELNs.  The imposed time step automatically identifies with the {\it observation time} for the statistics, thus eq.~(\ref{timestepcond}) introduces a conflict with eq.~(\ref{condsp}) since
\begin{eqnarray}
\label{eulercond}
\Delta t_s \ll \frac{1}{\Gamma_c} & \Rightarrow\!\Leftarrow &  \Delta t_s \gg \frac{1}{\Gamma_c} \, .
\end{eqnarray}

The inherent contradiction between the two requirements clearly indicates the existence of a numerical inconsistency.  For a concrete example, we focus on a laser with $\beta = 10^{-1}$; a full investigation for $\beta \in [10^{-6}, 1]$ (i.e., macro- to nano-laser) will be presented elsewhere~\cite{inprep}. 

\begin{figure}[htbp]
  \centering
  \includegraphics[width=0.95\linewidth,clip=true,trim= 10 40 10 70]{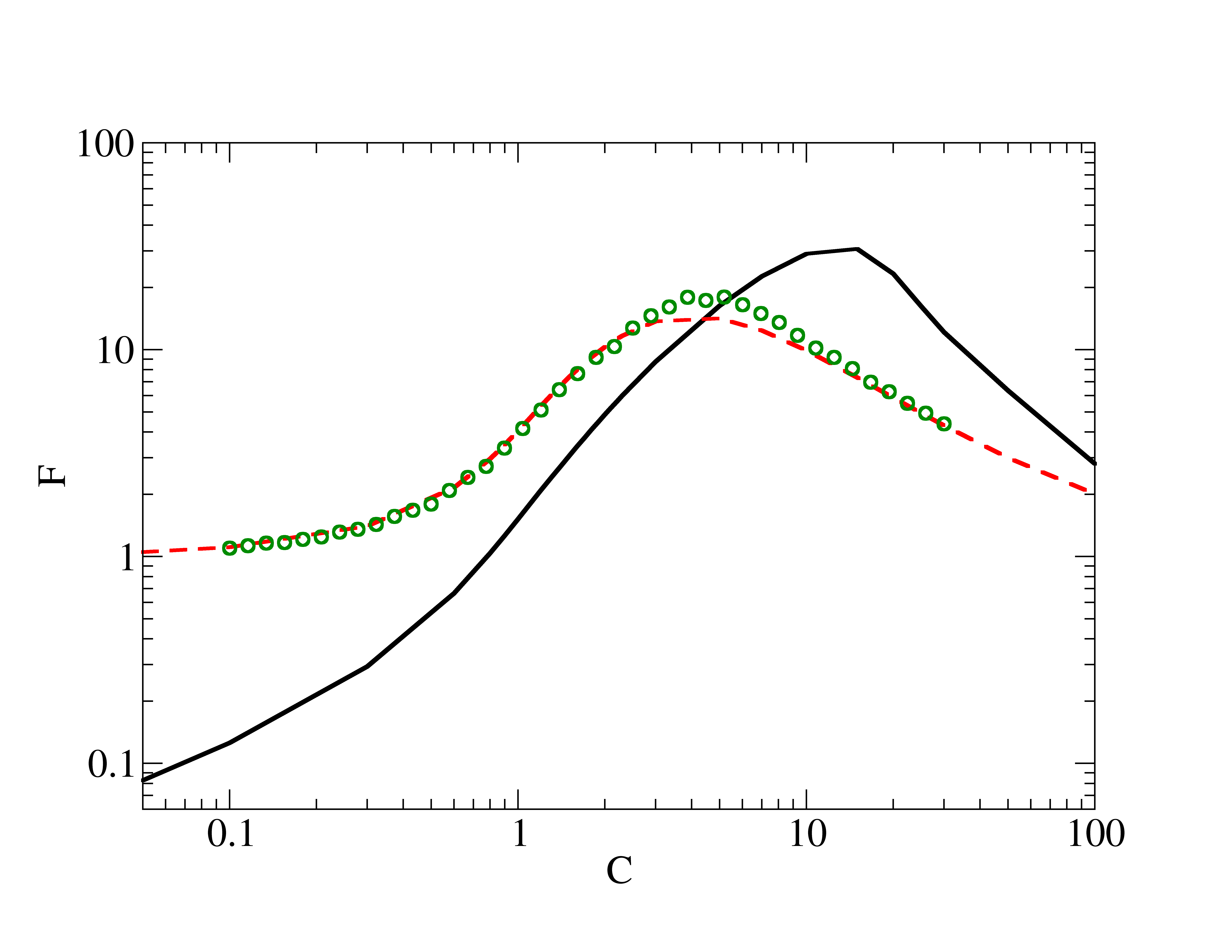}
\caption{Fano factor computed for $\beta = 10^{-1}$ (solid, black online) by numerically integrating the RELNs for $\gamma = 10^9 s^{-1}$, $\Gamma_c = 10^{11} s^{-1}$, and $\Delta t_s = 10^{-14} s$.  The dashed (red online) curve displays the analytical prediction coming from the RELNs and computed through a small-signal treatment (eq.~(\ref{anfano}--\ref{params})), while the open symbols correspond to the numerical predictions obtained from the stochastic simulator.} 
\label{fano} 
\end{figure}

Fulfilling the conditions outlined in~\cite{Lippi2018} for a correct integration of the RELNs, we obtain the numerical estimate of the Fano factor $F$~\cite{Rice1994}:
\begin{eqnarray}
\label{anfano}
F  =  \frac{\langle n^2 \rangle - \langle n \rangle^2}{\langle n \rangle} = \frac{\langle \Delta n^2 \rangle}{\langle n \rangle}
\end{eqnarray}
shown by the continuous line (black online) in Fig.~\ref{fano}.  We compare this result to the analytical prediction obtained through a small-signal analysis \cite{Mork2018} of the same RELNs, represented by the dashed curve (red online) in Fig.~\ref{fano}.  Neglecting the Langevin noise force in the carrier rate equation -- a good approximation in a wide pump range --,
the variance of the photon number takes the simple form
\begin{eqnarray}
\label{photonvariance}
\langle \Delta n^2 \rangle  \approx  \frac{\gamma_p \overline{n} (\overline{n} + 1)}{\gamma_p + \gamma_e} \left( 1+ \frac{\gamma_e^2}{\omega_{ro}^2 + \gamma_e \gamma_p} \right) \, , \\
\label{params}
\gamma_e  \equiv  \gamma \left( 1 + \beta \overline{n} \right) \,  , 
\gamma_p =  \frac{\Gamma_c}{\overline{n} + 1} \,\, ,
\omega_{ro}^2  =  \beta \gamma \Gamma_c \overline{n} \,\, .
\end{eqnarray}
The discrepancy between the two curves is striking:  below threshold the numerical prediction lies more than one order of magnitude below the analytical one; the maximum of $F$ (typically associated with threshold crossing) occurs at a pump value five times larger than in the analytics; finally, there is no convergence between two two curves at large pump values.

One should have expected the two estimates to coincide -- up to, at most, small details -- since $F$ is computed, albeit with different techniques, from the same set of RELNs.  Yet this is not the case.  Anticipating on what follows, the analytical prediction (dashed red line) is correct.  The discrepancy is not limited to $F$ but also occurs for the Relative Intensity Noise (RIN) and the zero-delay second order autocorrelation ($g^{(2)}(0)$) and is observed, to various degrees, for {\bf all} lasers~\cite{inprep} ($\beta \in [10^{-6}, 1]$).

While eq.~(\ref{eulercond}) clearly explains the failure of convergence of the numerical solution below threshold (and for increasing $R$ until the contradiction (\ref{eulercond}) is resolved) the disagreement at large pump values has a different origin, which we examine after offering an independent confirmation of the validity of the analytical result.

A better representation of the stochastic dynamical evolution can be obtained with a different numerical approach based on the Stochastic Laser Simulator (SLS)~\cite{Puccioni2015} where all physical processes (pumping the excited state, spontaneous and stimulated relaxation into the lasing mode, relaxation into other modes and non-radiative relaxations, as well as the transmission of photons through the cavity outcoupler) are described in a semiclassical fashion~\cite{Einstein1917} as probabilistic processes controlled by Poisson statistics.  Thus, the SLS amounts to {\it observing} the establishment of coherent emission through a sequence of probabilistic (poissonian) physical steps, defined through a recurrence relation.  Hence, no additional statistical hypotheses are introduced and results are realistic, within the bounds of a semiclassical description~\cite{Puccioni2015}.  Comparison with experimental observations has proven the validity of the SLS predictions, down to unexpected details~\cite{Wang2015}. 

\begin{figure}[htbp]
  \centering
  \includegraphics[width=\linewidth,clip=true,trim= 10 40 10 70]{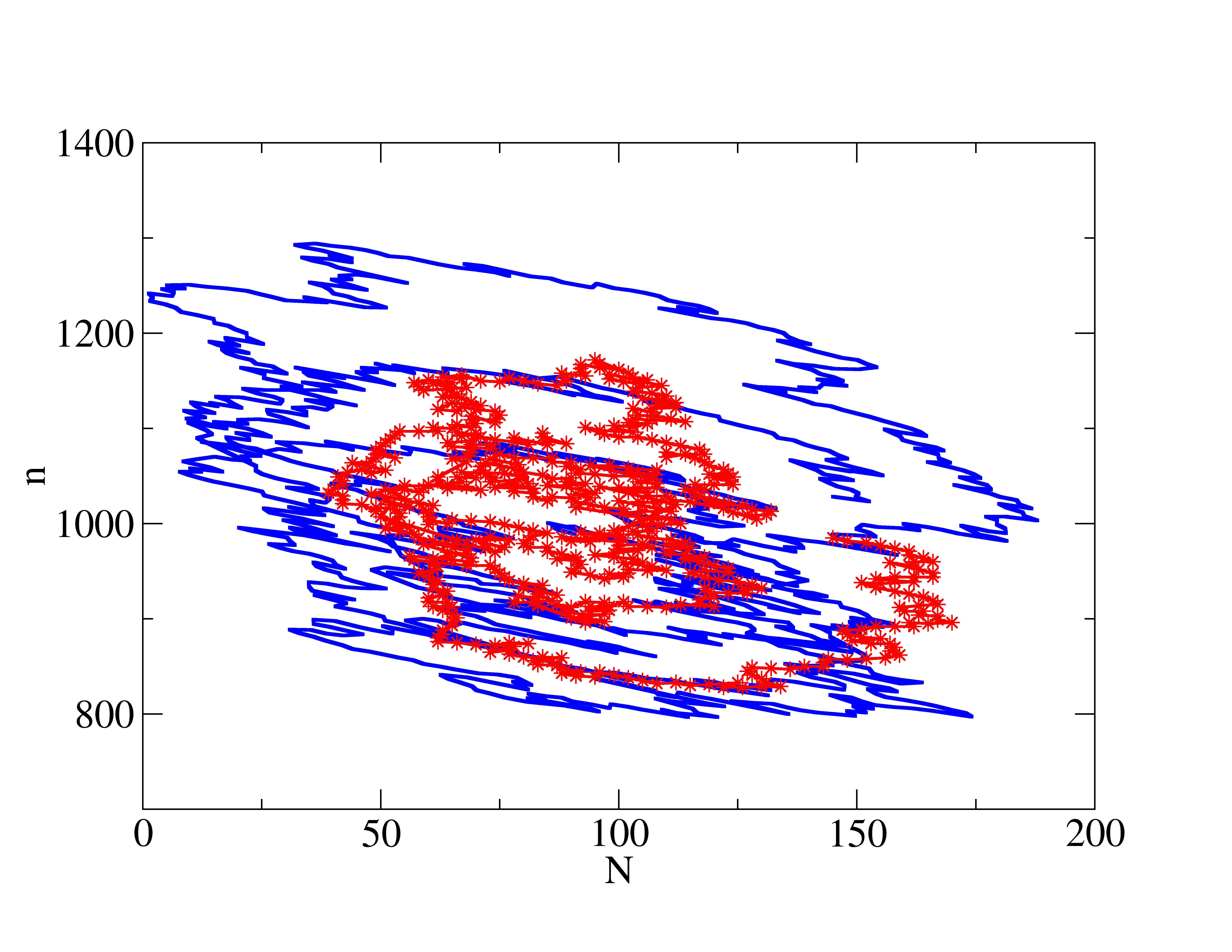}
\caption{Phase space representation of individual trajectories (photon number $n$ vs, carrier number $N$) computed by the RELN (solid curve, blue online) and by the SLS (symbols, red online) for $R = 10
R_{th}$  and $\beta = 10^{-1}$.  For the RELNs $\Delta t_s = 10^{-14} s$, while for the SLS $\Delta t \approx 10^{-15}$ (automatically adjusted to meet the requirements of the Poisson pumping process~\cite{Puccioni2015}).  For the SLS the photon number is computed ``inside'' the cavity to match the RELNs.  To avoid cluttering the figure with excessive details the data are smoothed over 7 consecutive points before plotting.} 
\label{traj} 
\end{figure}

Comparison between the analytical Fano factor (Fig.~\ref{fano}, dashed -- red -- line) and the predictions of the SLS (Fig.~\ref{fano}, open dots) supports the statement in favour of the validity of the analytics over the numerical integration of the RELNs: the excellent quantitative agreement over most of the explored range of pump values, both below and above threshold, is obtained with two entirely different approaches and without free parameters.  Only near the maximum Fano value a small discrepancy appears (about 20\% in amplitude and 7\% in position of the maximum), most likely due to having neglected the carrier noise contribution in the analytical derivation of eq.~(\ref{photonvariance}).  Excellent agreement was also observed in~\cite{Mork2018} for a system with a finite number of emitters.

Why should the analytical predictions of the RELNs be correct even for {\it large} $\beta$ lasers (small devices) while the numerics are patently wrong, even for smaller $\beta$ values?  Although not giving a mathematical proof for this observation, we can offer the following remarks which support it.  Analytical calculations do not impose timescales over which the computations are performed:  the analytical photon number variance is computed in the Fourier domain considering white noise, thus the temporal filtering imposed by $\Delta t_s$ in the numerical integration of the RELNs is not present.  Low-frequency components, excluded by eq.~(\ref{timestepcond}), allow in the analytics for sufficiently long {\it observation times}, solving the contradiction inherent in eq.~(\ref{timestepcond}).  Thus, the absence of constraints in the calculation technique removes the obstacles which prevent the numerics from providing meaningful results until threshold and beyond (until eqs.~(\ref{condsp},\ref{timestepcond}) are no longer in conflict).

At sufficiently large pump values one would expect the discrepancy between statistical and numerical integration to be lifted, thanks to the large number of photon emission processes which (assuming ergodicity) allow for a shorter {\it observation time}, compatible with the numerics.  Since Fig.~\ref{fano} shows this not to be the case, we turn to analyzing the temporal trajectories.  Fig.~\ref{traj} shows two trajectories computed ten times above threshold from the RELNs (solid, blue online) and from the SLS (symbols, red online), respectively.  The phase space clearly shows a meandering around the fixed point $(\overline{N} = 100, \overline{n} = 1000)$ in both cases, however RELN predictions are typically 10--20\% farther away from the fixed point.  The resulting larger deviations produce a larger variance and Fano factor.  

 For $C \gtrapprox 4$, two causes contribute to numerical estimates of the Fano factor being larger than the analytical ones.  The spectral filtering imposed by the time step $\Delta t_s$ numerically averages out all high-frequency noise components.  Such components, which correspond to time steps smaller than $\Delta t_s$, would contribute independent noise events which should quadratically add.  
Their removal leaves behind the low-frequency contributions  (thus larger, because of the longer time step) which lead to more substantial deviations from the actual trajectory.  The second element contributing to the distorted predictions is related to the structure of the phase space on which the trajectory of Fig.~\ref{traj} moves.  Not unexpectedly,
the {\it landscape} which describes the effective potential in phase space has  features which depend on the distance from the fixed point:  the local slope changes as one moves farther away from the steady-state solution.  Thus, trajectories evolving at different distances from the fixed point feel different gradients and develop accordingly.

\begin{figure}[htbp]
  \centering
  \includegraphics[width=\linewidth,clip=true,trim= 0 0 0 0]{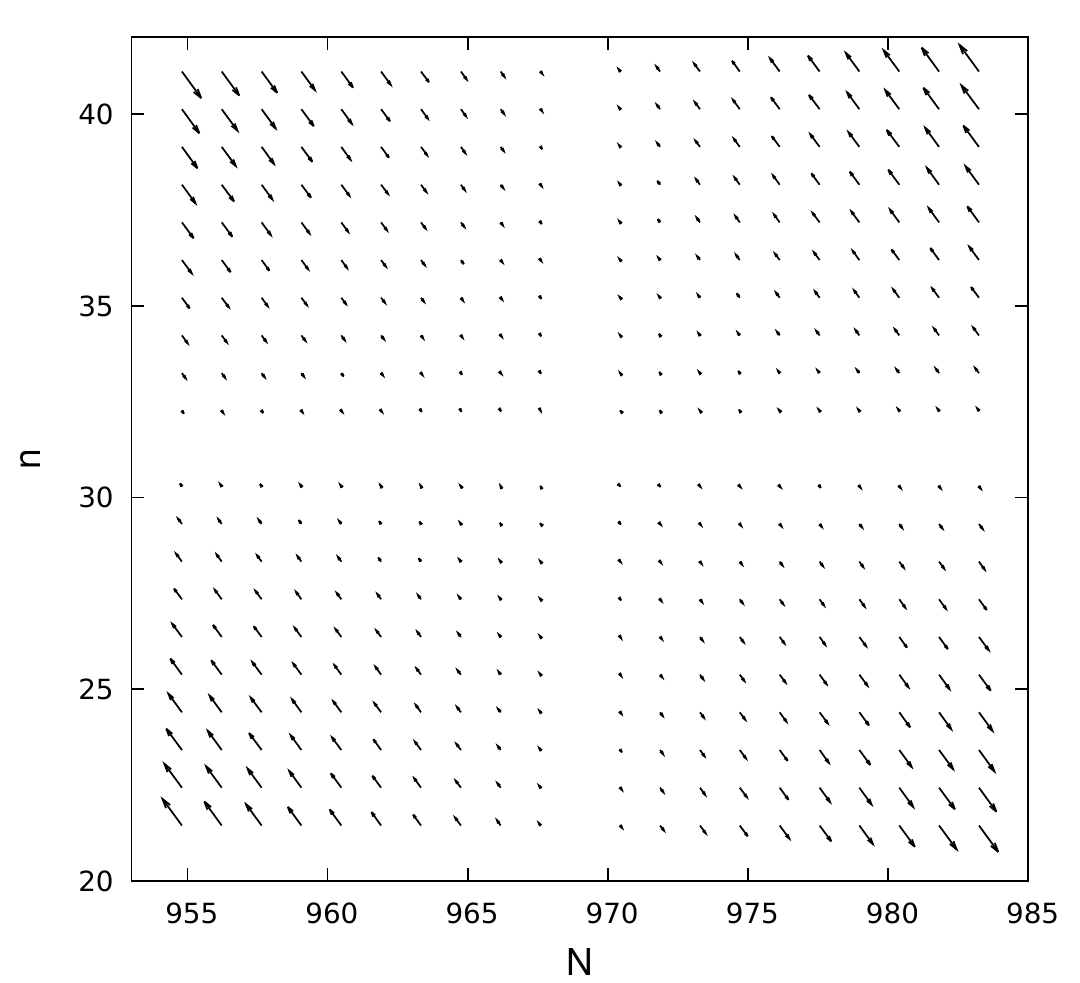}
\caption{
Vector field difference $\Delta \vec{v}$ around the fixed point for $R=4 R_{th}$.
}
\label{vecfield} 
\end{figure}

The overall (deterministic) structure of the phase space around the fixed point $\overline{P} = (\overline{N}, \overline{n})$ corresponds to a spiral (at the origin of laser relaxation oscillations, recognizable in Fig.~\ref{traj}) which ends into the stationary solution.  The vector field, depicting the local gradient for each $(N,n)$ point, is also a spiralling field~\cite{Lippi2000}.  The actual stochastic trajectory moves on top of this vector field, thus giving rise to the noisy, non-converging meandering powered by continuous fluctuations (Fig.~\ref{traj}).  The lack of linear scaling in the vector field -- as a function of the distance from the fixed point --, probed by the larger fluctuations of the Langevin approach, is the origin of the larger estimates in the Fano factor.  

Defining the vector field $d\vec{v}$ and its difference $\Delta \vec{v}$ at any point $P$ in the neighbourhood of the fixed point $\overline{P}$
\begin{eqnarray}
d\vec{v} = \left( \begin{array}{c} dN \\ dn \end{array} \right), & \forall & P = \left( \begin{array}{c} N \\ n \end{array} \right) \, ,\\
\Delta \vec{v} = d\vec{v}(\overline{N}+\eta_N,\overline{n}+\eta_n) & - & \alpha \,  d\vec{v}(\overline{N}+\epsilon_N,\overline{n}+\epsilon_n) \, , \\
\eta_j & = & \alpha \epsilon_j \, ,
\end{eqnarray}
permits the exploration of the vector field nonlinearity by comparing its structure in a small neighbourhood $\epsilon$ to that in a larger field $\eta$, where the vectors of coordinates $\epsilon_N$ and $\epsilon_n$ (thus $\eta_N$ and $\eta_n$) define the matrices of points on which $\Delta \vec{v}$ is computed.  Fig.~\ref{vecfield} shows $\Delta \vec{v}$ around the fixed point ($R = 4 R_{th}$) for $\alpha = 100$.
Depending on the quadrant in which the representative point is instantaneously located, the nonlinearity in the vector field causes either a deviation from top left to bottom right (2$^{\rm nd}$ and 4$^{\rm th}$ quadrants) or in the opposite direction leading away from the fixed point.  A careful examination shows that only the second quadrant gives an additional convergence towards the fixed point (i.e., a stronger curvature), while the other two quadrants either lead directly away from it (fourth quadrant) amplifying the deviation, or push the point diagonally away, resulting in a (smaller) amplification of the deviation.  

The combination of the nonlinearity of the phase space and the large excursions resulting from the numerical Langevin approach gives rise to the increased fluctuations in the full range of pump values where the RELNs predict larger values of $F$ (Fig.~\ref{fano}).

\begin{figure}[htbp]
  \centering
  \includegraphics[width=\linewidth,clip=true,trim= 0 0 10 0]{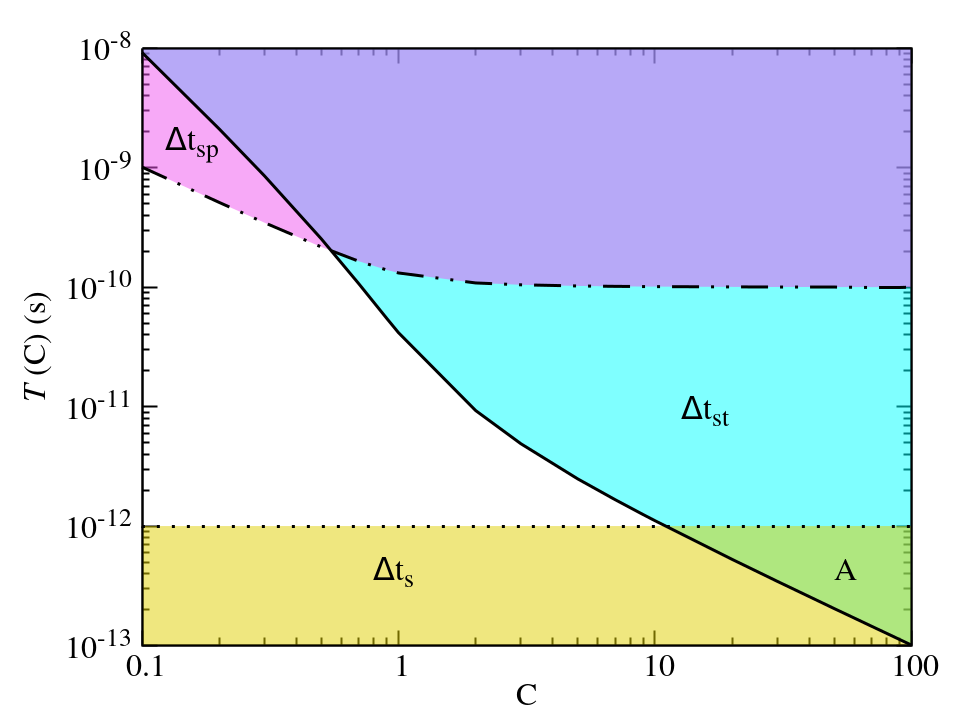}
\caption{
Compatibility of the various inequalities as a function of (normalized) pump, $C$.  The lines are traced setting the numerical coefficient equal to 10 in the right-hand-side of eqs.~(\ref{spontcond}) -- spontaneous emission, double-dotted--dashed line --, of eq.~(\ref{stimcond}) -- stimulated emission, solid line --,  and to 0.1 in eq.~(\ref{timestepcond}) -- time step condition, dotted line.  
}
\label{conditions} 
\end{figure}

An overview of the full restrictions imposed by the statistics and the numerical constraints (eqs.~(\ref{spontcond},\ref{stimcond},\ref{timestepcond}), collectively identified as $\mathcal{T}(C)$ in the following) further explains the differences between analytical and numerical curves (Fig.~\ref{fano}).  Fig.~\ref{conditions} graphically illustrates the inequalities of $\mathcal{T}(C)$ by separating the plane into regions where each individual condition is satisfied (coloured areas -- grey in print -- identified by the corresponding time step from eqs.~(\ref{spontcond},\ref{stimcond},\ref{timestepcond})).  
The only region fulfilling ``all''  conditions is marked by the letter $A$, where we do not consider inequality~(\ref{spontcond}), due to the negligible number of spontaneous events far above threshold ($R \ge 10 R_{th}$).  For all other pump values, one or more of the conditions is violated to a different degree, thus explaining the discrepancy between analytical and numerical analysis based on RELNs.  It is important to remark that the value chosen for the maximum time step $\Delta t_s$ ($10^{-12} s$) in  this figure is an upper estimate, which artificially enhances some spectral components and that better quantitative results may be obtained with smaller values (as discussed in~\cite{Lippi2018}, where a better choice is shown to be $\Delta t_s = 10^{-14} s$).

In conclusion, analytical predictions obtained from the Rate Equations with Langevin Noise and numerical results obtained from the Stochastic Laser Simulator support each other and offer two alternative approaches for the investigation of the noise properties in small--scale lasers (and possibly in other quantum--optical problems).  On the other hand, numerical approaches based directly on the  Rate Equations with Langevin noise are fraught with difficulties due to conflicting demands on the numerical time step and nonlinearities in phase space.

GLL is grateful to L. Gil for frequent discussions and very valuable advice.  JM acknowledges financial support from Villum Fonden via the NATEC Centre (Grant No. 8692).

\end{document}